# QoT Assessment of the Optical Spectrum as a Service in Disaggregated Network Scenarios

KAIDA KAEVAL[1,2,*], TOBIAS FEHENBERGER[1], JIM ZOU[1], SANDER LARS JANSEN[1], KLAUS GROBE[1], HELMUT GRIESSER[1], JÖRG-PETER ELBERS[1], MARKO TIKAS[3], GERT JERVAN[2]

[1] *ADVA Optical Networking SE, Martinsried, Germany*
[2] *Tallinn University of Technology, Tallinn, Estonia*
[3] *Tele2 Estonia, Tallinn, Estonia*
*\*Corresponding author: kkaeval@adva.com*

**The potential to operate third-party terminals over multi-domain transparent optical networks attracts operators and customers to implement Optical Spectrum as a Service (OSaaS). As infrastructure information cannot always be shared with OSaaS end customers, alternatives to off-line Quality of Transmission (QoT) estimation tools are required to assess the performance of the spectrum slot in order to estimate achievable throughput. In this paper, commercially available sliceable coherent transceivers are used to assess the Generalized Signal To Noise Ratio (GSNR) based QoT of the OSaaS in a live production network for both, narrow-band and wide-band OSaaS configurations. Extended channel probing based on symbol rate variability is combined with spectral sweeping and operation regime detection to characterize OSaaS implementations on 17 links with different underlying infrastructure configurations in order to maximize capacity and increase service margins in a low-margin operation regime. We achieve 0.05 dB estimation accuracy in GSNR for a wide-band spectrum services and 0.32 dB accuracy for narrow-band spectrum services. Based on the GSNR profile, spectral misalignment, spectral ripple and operation regime are detected and service margin improvements are demonstrated. Finally, we discuss the network optimization perspective based on acquired data from channel probing and propose use-cases for continuous channel probing in transparent optical networks.**

## 1. INTRODUCTION

With the high OSNR tolerance, automatic impairment mitigation and ultra-long reach of modern coherent transceivers, operators are eager to implement Optical Spectrum as a Service (OSaaS) [1, 2] in transparent optical networks. Unique for this service model, in OSaaS, the optical transceivers are owned and controlled by the service customer, whereas the open line system (OLS) is controlled by the network operator, creating the disaggregated networking environment in dense wavelength division multiplexing (DWDM) networks.

In essence, OSaaS is as a transparent lightpath connecting two endpoints in a single or multi-domain optical network, capable of carrying a single wavelength or multiple carriers over a predetermined spectrum. It is completely independent from the underlying infrastructure and can be applied to both, new flex-grid and legacy fixed-grid systems, as long as the end-to-end optical spectrum is available for the service. This allows transparent connections stretching thousands of kilometres between the terminal units, traversing multiple systems and operator domains, providing significant cost savings from both, investment and operational cost perspective.

In this paper, we refer to the spectral slot or a lightpath allocated for the OSaaS as a Media Channel (MC) as per ITU-T Recommendation G.807[3] and the wavelength service(s) inside it as Optical Tributary Signal (OTSi). The OTSi can be directly operated inside the MC, or a dedicated Network Media Channel (NMC) per OTSi can be configured. Depending on the access structures and ROADM functionalities on the OLS system, narrow-band or wide-band OSaaS can be configured. While fixed-grid access structures allow only narrow-band OSaaS configurations, typically used for "alien wavelength" services, systems with colourless access provide more options. Here, multiple adjacent narrow-band MCs or a single wide-band MC can be configured to form a wide-band OSaaS service. As signal power from alien carriers increases the vulnerability of the OLS system, narrow-band OSaaS is simpler to control and operate. However, wide-band configurations are preferred, when operation with high symbol rate signals or multiple carriers is desired to reduce the possible bandwidth narrowing effect.

A general OSaaS set-up with third-party terminals over the single and multi-domain open line system (OLS) is presented as scenario 1 and 2 in Fig. 1. While the most common usage scenario for OSaaS is a single vendor/domain scenario, the most benefits from OSaaS are obtained in a multi-domain environment. A perfect example of the multi-domain environment is formed by the national DWDM backbones of the small European Union countries that are connected together. A currently typical solution for the international connectivity in this scenario would require signal regeneration at the country borders, which can be avoided with OSaaS.

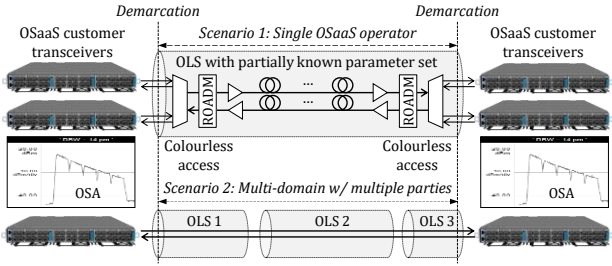

| Media channel characteristic | Value | Min | Max |
|---|---|---|---|
| Allocated bandwidth (GHz) | 400 | 193.75THz | 194.15THz |
| Total allowed power (dBm) | 9 | -1 | 9 |
| Allowed PSD (dBm/GHz) | -23 | -33 | -20 |
| OSNR$_{0.1nm}$ @ f$_{central}$ (dB) | 20 | 17 | 20 |
| GSNR @ f$_{central}$ (dB) | | | |
| Wavelength-dependent performance variations | Not available | | |
| Operation regime | | | |

Fig 1. Top: General OSaaS set-up with third-party terminals over the single and multi-domain open line system (OLS), Bottom: partially available parameter set per OLS

Contractually, OSaaS is described with the parameters listed at the bottom of Fig. 1. While the allocated bandwidth, total allowed power and power spectral density (PSD) define the input thresholds for the customer signal, the OSNR alone does not give a comprehensive overview about the achievable performance on the spectral slot. Essential characteristics to estimate the total achievable throughput of the OSaaS service, like GSNR, wavelength dependent performance variations and operation regime, are often not provided by the OLS operator.

To accurately estimate the highest achievable capacity per spectral slot and select the best suitable transceiver configuration out of thousands of possible configurations provided by the modern transceivers that still satisfy the minimum margin requirements [4], the user is required to account for all transmission impairments. This means accounting for any filtering penalties from various cascaded multiplexers, reconfigurable optical add/drop multiplexers (ROADMs) or even grating based dispersion compensation modules (DCGs) in fixed grid optical networks [5] to frequency-dependent spectral ripple caused by EDFA and Raman amplification profiles in wide-bandwidth flex-grid OSaaS scenarios. While the power-limited operation regime of the amplifiers sets the limits for the per-channel signal powers and optical power spectral density of the individual carriers, the launch powers set in the design phase of the link may not be the optimum for the selected route length. Multi-domain lightpath scenarios and any unpredictable time-varying changes in the QoT only add uncertainties to the quick yet reliable optimum performance-estimations exercise of the OSaaS. Furthermore, obtaining the knowledge about the OLS to accurately estimate the QoT in disaggregated networking scenarios may be a complicated task. As span lengths and precise system component parameters could reveal the achievable capacity and latency of the negotiated route, precise site locations along with system component data could increase the vulnerability to security related threats, making operators hesitating to share the data prior to contract signing.

To overcome the challenge, a channel probing method utilizing characterized coherent transceivers as a probing tool has been proposed to experimentally estimate the Generalized Signal to Noise Ratio (GSNR) of the lightpaths [5,6]. The usage of GSNR has proven to give reliable results in comparison with an offline GSNR based QoT estimation tool [7, 8] as well as on numerous field trials that introduce autonomous transceivers, combat routing challenges in infrastructure aware networking and estimate channel performance [9,10,11,12,13,14]. Furthermore, the channel probing method is applicable for both OSaaS usage scenarios in Fig. 1 and can be, with small modifications to the procedure, successfully implemented to precisely capture the spectrum performance regardless of the underlying infrastructure.

The aim of this paper is to assess the unavailable OSaaS performance metrics and accordingly optimize the transceiver configurations to maximize the achievable throughput on the spectral slot allocated for OSaaS. To this end, we present a comprehensive study of OSaaS performance assessments in field deployments in a black-box-scenario [9]. We extend our previous work [15,16] with additional routes and network configurations to cover a wide variety of OSaaS configurations in live networks and propose a channel probing toolkit, consisting of symbol rate variable extended channel probing, frequency sweep and operation regime detection. Considering both flex-grid and fixed-grid production network infrastructures, we experimentally demonstrate that the proposed toolkit is able to derive OSaaS performance characteristics for any link with narrow-band or wide-band OSaaS deployment. Finally, we discuss how the knowledge acquired from channel probing can lead to increased capacity and service margins in low-margin networking scenarios.

The structure of this paper is as follows. Section 2 provides an overview about the generic channel probing concept to be used as a primary tool to characterize the OSaaS spectrum. In Section 3, general system description and test-setup is introduced, following with the results and findings explained in Section 4. The results section presents the media channel characteristics captured according to the adjusted channel probing parameters, followed by the discussion with some additional optimization possibilities in Section 5.

## 2. CHANNEL PROBING

Channel probing in transparent optical networks is not a new concept. Probes called active, supervisory or dummy lightpaths were used already in the early 2000's to estimate the link availability and performance in transparent optical networks [17,18,19]. These lightpaths used intensity-modulated direct-detection (IMDD) transceivers as the probing light and were commissioned to extract availability and pre-FEC bit error ratio (BER) performance data from the network link. However, the perspective of link performance estimation precision, accuracy and usage-scenarios in optical networks has been significantly widened with the concept to use characterized DSP-based modern sliceable transceivers as the probing light.

The very first attempt to use a back-to-back characterized coherent transceiver to capture the link performance was performed by Torrengo et al in 2011 [20], when they introduced the first lab experiment to verify the predecessor of the GN model [21]. The GN model by Poggiolini [22] poses that both components, the linear amplified spontaneous emission (ASE) noise and nonlinear impairments (NLI) contributing to the Generalized OSNR would have a Gaussian distribution in case of non-dispersion compensated spans. This led to a symbol rate independent GSNR to became a widely used optical link QoT metric, when the model was verified also through simulations [23,24]. Starting from 2016, the topic has gained attention in optical submarine networks as part of the implementations of disaggregated networking scenarios over the trans-Atlantic and trans-Pacific links, introducing GSNR as the primary link characteristic [25,26] and proposing instructions for transceiver characterizations to perform accurate GSNR measurements. Since 2017, characterized coherent transceivers to estimate the link GSNR have been used in many studies, from optical link characterization [27,28,29,25] and network automation [13,30] to linear OSNR estimations [10, 31, 32], but it was only after the more intense insights to GSNR-based link characterization through characterized transceivers, provided in [5], when the topic gained real momentum.

### A. General channel probing concept

To evaluate the actual channel performance of a lightpath at hand, a characterized probing-light with a fixed modulation format and symbol rate is inserted into the network in the corresponding channel location, and the pre-FEC BER estimation of the receiver, converted into a Q-value, is used to estimate the respective effective Generalized Optical Signal to Noise Ratio (GOSNR). The estimated GOSNR considers all optical distortions that impact the optical signal, including ASE noise, nonlinear distortions, as well as any transceiver impairments. This value is then normalized to the symbol rate of the Probing Light Transceiver (PLT) signal to obtain the estimated GSNR of the link ($GSNR_{est,link}$). Fig. 2 explains the general channel probing process.

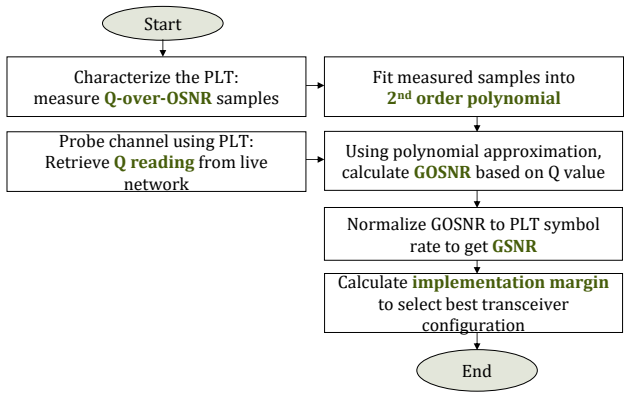

Fig 2. General channel probing concept

The accuracy of the $GSNR_{est,link}$ is defined as the difference between the estimated GSNR and the real GSNR of the link. While the characterization errors, OLS equalization and PLT data reading errors contribute to systematic error, random errors of the $GSNR_{est,link}$ are caused by performance fluctuations in time. To increase the accuracy, multiple $GSNR_{est,link}$ estimations from different PLT configurations used in a constant power spectral density mode, as per [15] can be captured. This allows to average out the errors from probing procedure and use mean value from all estimations as a final estimated $GSNR_{est,link}$.

GSNR estimations received from the channel probing form the original probing results for our spectrum QoT assessment and optimum transceiver configuration selection.

### B. Accuracy of the method

To select the best possible transceiver configuration for the probed link, the GSNR implementation margin ($GSNR_{margin}$) is estimated for any tested modulation format/symbol rate configuration. To do so, typically required GSNR ($GSNR_{req}$) per configuration, available from the system specification documentation is subtracted from the estimated link GSNR ($GSNR_{est,link}$):

$$GSNR_{margin} = GSNR_{est,link} - GSNR_{req} \quad (1)$$

All calculations resulting in a positive GSNR implementation margin are expected to work over the probed link and all calculations resulting in a negative GSNR implementation margin, not to work.

As the theoretical GSNR calculations are often not available in OSaaS, the verification of the estimated $GSNR_{est,link}$ accuracy has to be also addressed experimentally. To verify the accuracy of the estimated $GSNR_{est,link}$, near zero-margin characterized PLT configurations can be used as verification signals by comparing the estimated $GSNR_{margin}$ from equation (1) to the actual performance of the PLT configuration over the tested link. Any contradiction between the estimation and real signal condition on the tested network link would indicate a false $GSNR_{est,link}$ estimation. The accuracy of the $GSNR_{est,link}$ is specified through the $GSNR_{margin}$, and is the absolute value of the highest false $GSNR_{margin}$ prediction. To increase the accuracy of the $GSNR_{margin}$ estimation, multiple $GSNR_{est,link}$ estimations per link can be captured to reduce random errors from the channel probing procedure and mean value from all estimations can be used as the final $GSNR_{est,link}$ estimation in equation (1).

According to [5], $GSNR_{est,link}$ obtained from the probing measurement can be defined as the ratio between the power of useful signal divided by the sum of the powers of all noise sources – such as ASE ($SNR_{ASE}$), NLI ($SNR_{NLI}$) and additional noise arising from the specific PLT unit modem technology used ($SNR_{modem}$) - evaluated wholly in the signal bandwidth and expressed as:

$$\frac{1}{GSNR_{est,link}} = \frac{1}{SNR_{ASE}} + \frac{1}{SNR_{NLI}} + \frac{1}{SNR_{modem}} \quad (2)$$

Therefore, without characterizing propagation associated penalties such as CD, PMD and PDL associated with the test, the real GSNR of the link, consisting of the $SNR_{ASE}$ and $SNR_{NLI}$ component only, is not available. However, $GSNR_{est,link}$ estimations are valid for the $GSNR_{margin}$ estimations for the same modem type. In [5, 33], the characterization process of the PLT unit is explained and in [10, 33], the standard deviation and accuracy for the different pre-FEC BER readings and its impact on the achievable throughput estimation dependent on a performance variability of different PLT units is explained. These references conclude that a better GSNR estimation accuracy is achieved at lower Q values since the transceiver performance variation is much smaller in that region.

Leaving aside the PLT unit dependent penalties, equation (2) is valid in optical systems, where fibre propagation impairments caused by the interplay of loss, chromatic dispersion and Kerr nonlinearity can be approximated as additive white Gaussian noise (AWGN) [22]. This raises the question of GSNR metric reliability in short spans, dispersion compensated spans and spans experiencing high nonlinear penalty. Therefore, it is important to note that the estimated $GSNR_{est,link}$ from our channel probing is not the actual, real GSNR of the link, but its projection/approximation in the linear space. Our probing results from 17 tested links show that the estimated $GSNR_{est,link}$ approximation is well usable for $GSNR_{margin}$ estimations. We attribute this to the fact, that the GN-model is meant to be used in the linear domain, where also the Q-over-OSNR characterization of the PLT unit is performed. Therefore, any measured Q value from the real-life networks is compared to the polynomial fitting from the linear domain, projecting all the link induced impairments into linear domain, too. As also the typical, required GSNR from the system specifications is only defined for linear working regime, the $GSNR_{margin}$ estimations lead us to similarly accurate estimations in each of the tested network scenarios (including short and dispersion compensated links).

### C. GSNR estimation validity

In general, OSaaS characterization is performed once during the service handover. Although channel probing provides reliable end-to-end GSNR estimations, the estimation results are only valid for the infrastructure configuration and network load during the probing activity. There are two options to accommodate the future degradations from possible daily fluctuations, aging effects and network load/condition changes: a) incorporate additional margins as per [4] to the initial estimated $GSNR_{margin}$ to increase the robustness of the OSaaS live traffic or b) implement continuous probing. While the first is a fixed measure and only protects the OSaaS user within the degradation range set by margin, the later provides continuous performance estimations from the network. When data from continuous probing is used to train the machine learning (ML) algorithm, performance changes in OLS can be predicted and notifications can be sent to OSaaS user in case of abnormal changes.

Any major change in the network (including power distribution change between channels, changes causing increased filtering penalty, reroute, etc) should be carried out during the maintenance window and followed by a new OSaaS characterization. Depending on the width of the spectrum slot, full characterization incorporating extended channel probing, frequency sweep and operation regime detection can take up to an hour for a 100-GHz OSaaS service. The time required is primarily dependent on the number of PLT configurations used and stabilization times of the commercial transceivers. The set of PLT configurations can be customized according to link parameters.

## 3. GENERAL SYSTEM AND TEST SET-UP DESCRIPTION

To demonstrate the feasibility of the channel probing method in field deployments, we tested 17 different links in Tele2 Estonia's live networks. Twelve (12) links with lengths from 3 km to 1302 km were tested in a 10-Gbit/s optimized fixed-filter dispersion compensated ROADM based regional-haul network (routes A, B and C). The OSaaS service is configured with the 100-GHz nominal slot size in this network, making this a narrow-band OSaaS service, as the signals were added and dropped by arrayed waveguide grating (AWG) filters. Route A uses dispersion compensating fibres (DCFs) for optical dispersion compensation, whereas route B and C use a mix of DCFs and dispersion compensating gratings (DCGs). The central frequency of 193.2 THz was used on all of the links. Five links with lengths from 1016 km to 5738 km were tested in a pan-European coherent optimized flex-grid ROADM based long-haul dispersion compensation module (DCM) free network with colourless access architecture (route LH). The nominal slot size in this network is 50 GHz and a 400 GHz wide media channel with the central frequency 193.95 THz was configured for the channel probing tests, illustrating the wide-band OSaaS scenario. The OSaaS user spectrum was added and dropped at a free terminal ROADM port using 8:1 splitter/combiner module in test-site without applying additional optical filtering.

The underlying fibre infrastructure for all links and both networks conforms to the ITU-T 652.D standard for standard single mode fibres (SSMF). To extend the transmission distance and to allow single-ended measurements at test-site, the spectrum services were looped back in the far end ROADMs. Link specific data is given in Table 1.

The test set-up and spectral assignments of the links are illustrated in Fig. 3. Spans with black colour refer to DCM-free network infrastructure, while grey and orange mark the links compensated via dispersion compensation fibres (DCF) or dispersion compensation gratings (DCG) modules, respectively. Link acronyms are combined from route name and looped length of the link, for example A-144, B-485, C-284 or LH-1016. The only channel routed over two different routes on the link is B+A-1302 that was first routed over route B and then, starting from B-822, over route A.

Table. 1: Link data

| Loop-back | Network | DCM data | Looped link length | No of spans |
|---|---|---|---|---|
| A-4 | Regional | DCM free | 4km | 2 |
| A-144 | Regional | DCF | 144km | 4 |
| A-241 | Regional | DCF | 241km | 6 |
| A-382 | Regional | DCF | 382km | 8 |
| A-652 | Regional | DCF | 652km | 12 |
| B-70 | Regional | DCF | 70km | 4 |
| B-485 | Regional | DCF+DCG | 485km | 8 |
| B*-621 | Regional | DCF+DCG | 621km | 10 |
| B-822 | Regional | DCF+DCG | 822km | 12 |
| B-1182 | Regional | DCF+DCG | 1182km | 16 |
| B+2-1302 | Regional | DCF+DCG | 1302km | 18 |
| C-284 | Regional | DCF+DCG | 284km | 4 |
| LH-1016 | Regional | DCM free | 1016 km | 14 |
| LH-1792 | Long-haul | DCM free | 1792 km | 24 |
| LH-2943 | Long-haul | DCM free | 1792 km | 36 |
| LH-3751 | Long-haul | DCM free | 3751 km | 48 |
| LH-5738 | Long-haul | DCM free | 1792 km | 74 |

The probing unit was implemented on reconfigurable TeraFlex transceiver from ADVA, providing 100 Gbit/s to 600 Gbit/s capacity per carrier by adjusting the modulation format and symbol rate. All tests were run with reduced power levels compared to the allowed maximum power levels per media channel, to avoid any impact on the live channels. Therefore the results do not allow to derive the end-of-life network capacity capabilities, which is business-critical information for the network operator.

The field networks with different sets of infrastructure components, like multiplexers, amplifiers, boosters, and ROADMs, set a good playground to demonstrate the capabilities of the channel probing toolkit in characterizing the spectral slot performance and optimizing the transceiver configuration, operation regime and transmission impairments regardless of the underlying network infrastructure and condition.

The probing exercises were carried out over a ten-month time-period between May 2020 and February 2021. For all estimated GSNR estimations in this work, a constant power spectral density was used.

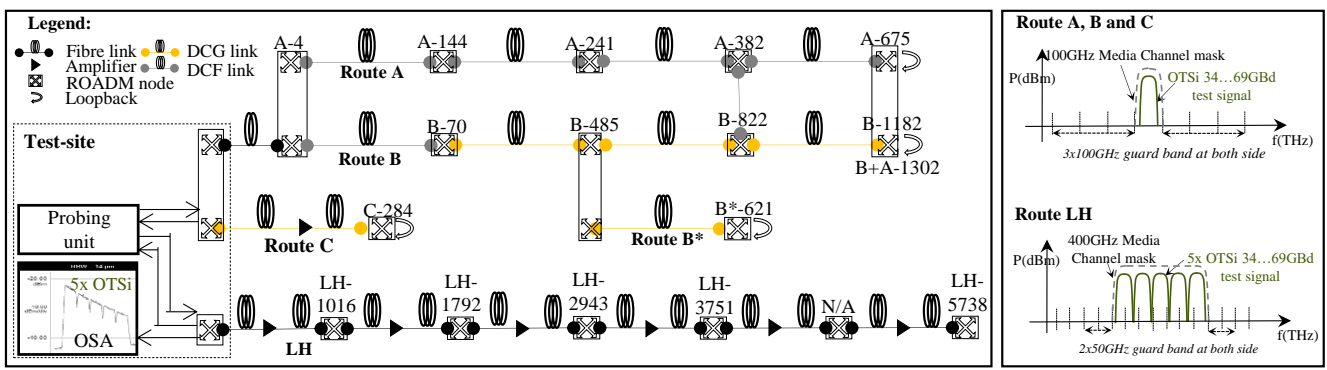

Fig. 3 General test set-up and spectral assignments on the routes

## 4. RESULTS

In this section we use channel probing method introduced in Section 2 to assess the missing parameters from Fig. 1 to fully characterize the spectrum slot performance and optimize the transceiver configuration to maximize the throughput. By adapting the symbol rate, frequency and power of the PLT configuration, we detect the possible filtering effects and estimate the GSNR of the spectrum slot, capture wavelength dependent performance differences within the OSaaS service and detect the operation regime.

### A. Symbol rate variable channel probing

As the generic channel probing method introduced in section 2 can provide reliable spectrum performance estimations only for the exact bandwidth covered by the PLT configuration in the same probed power regime, a single measurement may introduce a bias in $GSNR_{est,link}$ estimations in case of bandwidth limited OSaaS. To address this, we use eleven pre-characterized PLT configurations in power spectral density-based probing regime, as suggested in [15] to estimate the $GSNR_{est,link}$.

As the maximum achievable linear OSNR of the link is fixed due to design-based power levels, the non-linear impairments from the transmission media set the maximum achievable GOSNR on the link. However, optical links can introduce bandwidth narrowing due to ROADMs, multiplexers or grating based dispersion compensation modules. While the mild penalty from bandwidth narrowing can be compensated in the transceivers, any significant bandwidth limitation could cause the degradation on achievable performance.

In networks without bandwidth limitations, probing operated with constant PSD should return constant GSNR estimations regardless of the required bandwidth of the PLT configuration. Fig. 4 (a) presents the original probing results from the long-haul dispersion compensation free network with a colourless access architecture. The X-axis present the link length in kilometres and the Y-axis the estimated GSNR, obtained by using different PLT configurations. As visible, all measurement results regardless of the PLT configuration are concentrated within ±0.4 dB around the probable GSNR of the link with no differentiation from the used symbol rate or the modulation format of the PLT unit.

This sets a straight-forward way for $GSNR_{est,link}$ calculations based on the average estimated value from all of the working PLT configurations. Achieved $GSNR_{est,link}$ can be then further used to derive $GSNR_{margin}$ as per equation (1). Fig. 4 (b) presents the achievable accuracy of the $GSNR_{margin}$ estimation based on the averaged $GSNR_{est,link}$ estimation from all working PLT configurations in long-haul network. Cross marks accompanying the near-zero margin estimation indicates the false estimation, where the verification signal condition as per measurement did not agree the estimated signal condition. Here, the X-axis presents the link length in kilometres and the Y-axis the estimated $GSNR_{margin}$ for different PLT configurations. Different marker styles refer to different symbol rates and line styles distinguish different modulation formats. The chart is zoomed in for the area near zero implementation margin, where the probability of false estimations is highest. Using symbol rate variable channel probing, 0.05 dB in $GSNR_{margin}$ accuracy in the long-haul network was achieved compared to previously reported 0.7 dB [34]. This can be attributed to multiple GSNR estimations for the same link provided by symbol rate variable probing, as it allows to average out the small time-dependent performance fluctuations in the network that cause different BER readings, and thus the variation in Q-value per each measurement.

Modern flex-grid networks with colourless access architectures are less prone to experience bandwidth limitations or these can be overcome by simple reconfigurations in the ROADMs. Bandwidth narrowing, however, is a problem in legacy fixed-filter systems. In case of significant bandwidth limitation, high symbol rate signals are subject

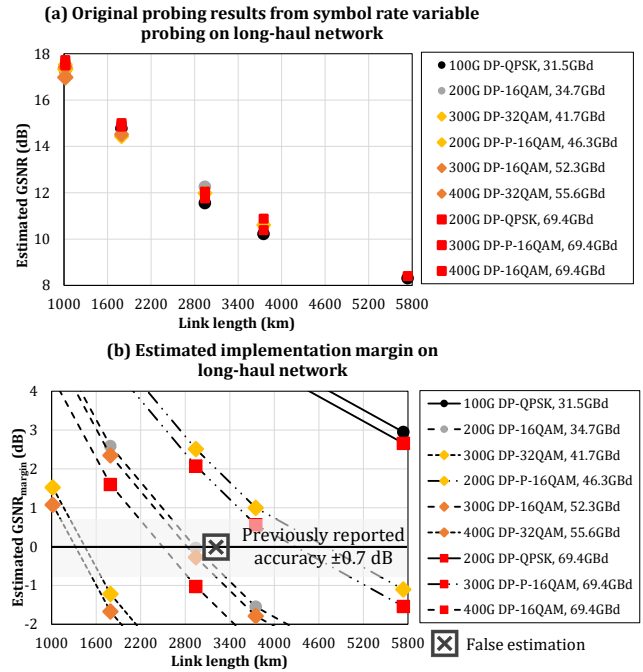

Fig. 4 (a) Results from symbol rate variable channel probing and (b) $GSNR_{margin}$ estimations for a long-haul network to stronger narrow band filtering and using a wideband probing signal results in underestimating the link GSNR that would be achievable for lower symbol rate signals. Using narrow band probing configurations would create the illusion of a link with a high GSNR not achievable for higher symbol rate signals.

This effect can be assessed with GSNR penalty, which is a systematic decrease in estimated GSNR, caused by the variation in PLT symbol rate, modulation format, frequency or power setting. As theoretical, calculated GSNR values for the link are often not available for the OSaaS end user, GSNR penalty for the specific configuration under interest can be calculated as a GSNR estimation difference between the highest performing PLT configuration and the configuration under interest, measured on the same transmission link. To verify the symbol rate dependent changes in GSNR estimations caused by bandwidth limitation and eliminate the GSNR estimation differences due to non-linear behaviour, back-to-back measurements (with 0-km fibre length) to characterize the filtering penalty were carried out in the lab, emulating the first link on route A. This link includes two ROADM modules and three arrayed waveguide grating (AWG) based multiplexer modules and defines the minimum filtering penalty on all the links in the regional-haul network. Fig. 5 presents the results of the back-to-back (b-2-b) measurements for the symbol rate dependency of the estimated GSNR. The X-axis represent the symbol rate of the PLT configuration and the Y-axis present the estimated GSNR. While the estimated GSNR decreases up to 0.6 dB compared to the highest performing PLT configuration at 34.5 GBd for the symbol rates up to 55.5 GBd, PLT configurations with 69.4 GBd experience significantly higher GSNR penalty, reaching almost up to 2.5 dB. This behaviour can be explained with a bandwidth limitation penalty from the AWG based channel filter modules (3-dB bandwidth of 80 GHz) and ROADMs.

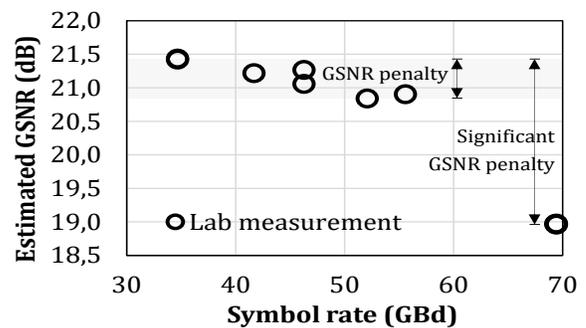

Fig. 5 Symbol rate variable probing results from b-2-b measurement

The probing results from the symbol rate variable channel probing in regional-haul network are presented in Fig. 6 (a). The X-axis presents the link length in kilometres and the Y-axis the estimated GSNR in dB. As expected, the link GSNR estimations by different PLT configurations are highly scattered for the legacy regional-haul links due to symbol rate dependent GSNR penalty caused by bandwidth narrowing. This leaves the average estimated GSNR value just as a statistical number for the bandwidth limited regional-haul network and further analyses on the original probing results are required to estimate the $GSNR_{est,link}$.

According to our results on Fig. 6 (a), PLT configurations with equal symbol rate experience similar penalties from the system and estimate the link GSNR with ±0.35 dB accuracy, regardless of the modulation type for up to 55.5 GBd signals, whereas a change to 69.4 GBd symbol rate in the PLT configuration resulted in a great variance in the estimated GSNR on live links. While the maximum variance between the estimated GSNR on links shorter than 500 km is below 0.7 dB for the probe settings between 31.5 and 46.3 GBd and below 1.1 dB for the probes up to 55.6 GBd, it quickly grows up to 7.1 dB for the probes including 69.4 GBd. For the links up to 1000 km, these figures are 0.7 dB, 1.3 dB and 7.5 dB, respectively. In addition to the bandwidth limitation penalty from the channel filter modules and ROADMs for the DCF based links, additional penalty from DCG modules (3-dB bandwidth of 60 GHz) is captured on the links including grating based dispersion compensation modules. For the two longest links in regional-haul network, only two PLT configurations between were working, having a GSNR penalty of 1.3 dB between the 31.5 and 46.3 GBd configurations. Fig. 6 (b) presents GSNR penalty for the PLT configurations that do not experience severe GSNR penalty from filtering. As visible from the figure, cascaded filtering from multiplexers, ROADMs and DCG modules decreases the effective bandwidth of the OSaaS on longer links, increasing the GSNR penalty even for the same PLT configurations.

In order to estimate the best possible transceiver configuration for the OSaaS offering at hand, it becomes essential to identify the highest usable symbol rate on the link – a symbol rate cap. To identify the symbol rate cap for the link, two criteria must be met: a) it must be acquired through working PLT configuration with the highest symbol rate and b) it does not experience significantly higher GSNR penalty compared to the other working PLT configurations on the link. According to our tested OSaaS services on Fig. 6 (b), four of the shortest links have a symbol rate cap 55.6 GBd. That is decreased to 52.3 GBd for links up to 1000 km and only 46.3 GBd for the longest two links. As visible from the Fig. 6 (b), the absolute values of the GSNR penalty itself are not consistent along the links nor important as long the two criteria for the symbol rate cap identification are met. All configurations exceeding the symbol rate cap should be removed from the pool of possible transceiver configurations before the $GSNR_{margin}$ calculation. In Fig. 6 (a), this means leaving out all high symbol rate configurations marked with red.

Then, the average estimated $GSNR_{est,link}$ can be calculated also for the bandwidth limited network, using the GSNR estimation results from PLT configurations with smaller and equal symbol rate than the symbol rate cap. The best transceiver configuration can be selected as per highest line rate from all possible configurations returning positive implementation margin as per equation (1).

Fig. 6 (c) presents the achievable accuracy of the $GSNR_{margin}$ estimation in legacy network together with the verification signal condition. The X-axis presents the link lengths in kilometres, and the Y-axis estimated $GSNR_{margin}$ in dB-s for the different PLT configurations. Different marker styles refer to different symbol rate regimes and line styles distinguish different modulation formats. The chart is zoomed in for the area near zero implementation margin and cross marks placed near the $GSNR_{margin}$ marker indicate the correctness of the estimation.

Following the identification of the symbol rate cap and applying this to original data from symbol rate variable channel probing, we achieved the selection of the best possible transceiver configuration with the estimation accuracy better than 0.10 dB for the links up to 822 km and 0.32 dB on the links up to 1302 km on the regional-haul network. The primary errors contributing to the false positive $GSNR_{margin}$ estimations were on the link B-1182 km, where the small number of working PLT configurations may have not contributed enough for the required averaging effect, to achieve more accurate $GSNR_{est,link}$ estimation. However, considering the margin values generally implemented for network robustness against aging and slow performance changes, the achieved result is an accurate measure of the network performance at a current moment.

### B. Channel probing with frequency sweep

To capture the possible performance differences over the wider spectral slots, generic channel probing procedure introduced in section 2 must be repeated with sweeping the central frequency over the available spectral slot with a suitably selected frequency step increments. For this, the frequency sweep exercise, similar to proposed in [35] was modified to OSaaS use-case, to detect the possible bandwidth limitations, frequency misalignments and spectral ripple/tilt on links A-241, C-284 and LH-1792.

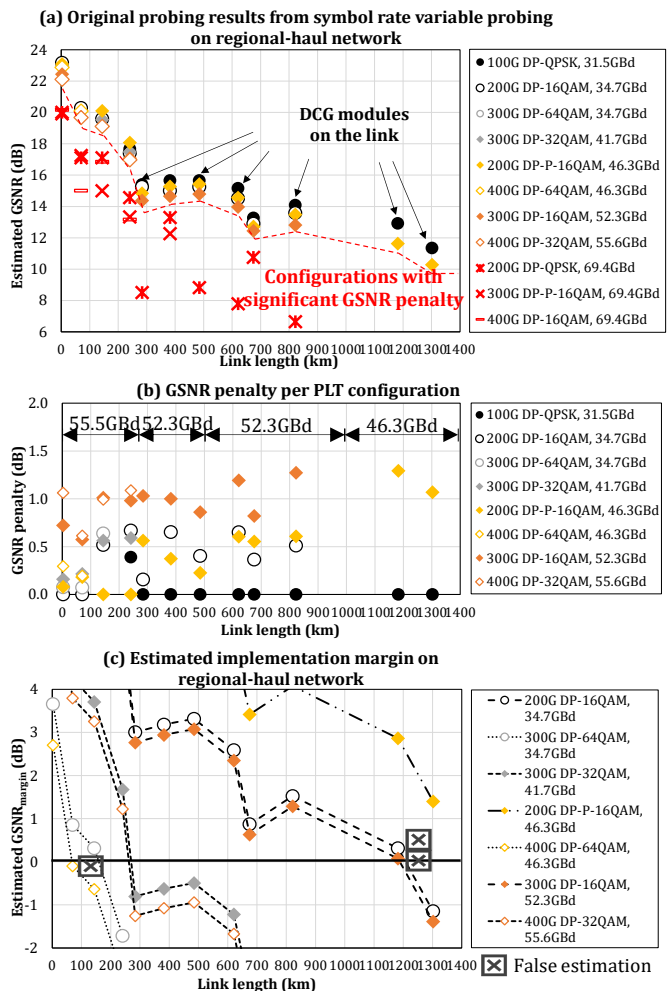

Fig, 6 (a) Results from symbol rate variable channel probing, (b) GSNR penalty and symbol rate cap and (c) $GSNR_{margin}$ estimations for a regional-haul network

A 100-GHz narrow-band spectrum service was investigated on routes A and C, whereas a 400-GHz wide-band service was used on route LH. Spectrum assignment for the tested routes are brought in Fig. 3. The channel performance on all routes was tested with reduced number of probes, utilizing only 200-Gbit/s 69-GBd DP-QPSK, 200-Gbit/s 46-GBd DP-P-16QAM, and 200-Gbit/s 34-GBd DP-16QAM probing signals with root raised cosine (RRC) spectral shape (roll-off factor r = 0.19). The DP-P-16QAM modulation format stands for a proprietary DP-8QAM, which is implemented using only partial constellation points from the standard constellation of the DP-16QAM modulation format. To obtain comparable results, constant power spectral density is maintained for all PLT configurations. This is possible when the DWDM line systems and their amplifiers operate in constant gain mode. The spectrum performance is evaluated by a GSNR, captured by sweeping in 6.25 GHz frequency steps over the allocated spectrum.

Fig. 7 shows the results of the channel probing on the routes A and C. While the nominal media channel width on route A and C is 100 GHz, the large predicted GSNR difference between the three modulation formats reveals that the usable effective optical bandwidth must be much smaller. Given that a 6.25-GHz mismatch of a 34-GBd signal already causes a noticeable penalty, we can conclude that the effective optical channel bandwidth is lower than 50 GHz. In Fig. 7 (a), up to 2 dB degradation in the estimated GSNR is observed for a 200-Gbit/s 34-GBd DP-16QAM signal configuration in case of 18.75 GHz offset. With the same offset, 200-Gbit/s 46-GBd DP-P-16QAM and 200-Gbit/s 69-GBd DP-QPSK signals are already experiencing outage as the attainable GSNR is insufficient.

The results in Fig. 7 (b) also indicate a misalignment of the nominal centre frequency on route C in addition to severe filtering penalty for the 200 Gbit/s 69-GBd DP-QPSK signal format. Without sweeping, this misalignment could have been left undiscovered and already a small fluctuation in the transceiver wavelength could cause a severe degradation of the service quality, especially with 46-GBd DP-P-16QAM modulation that generally requires higher margin, than a 200 Gbit/s 69-GBd DP-QPSK signal. To increase the robustness, the transceiver frequency should be fine-tuned to minimize frequency misalignment and to obtain the best spectrum performance.

Fig. 8 presents the GSNR variation over the LH-1792 link in a single 400 GHz media channel and a 5x75 GHz adjacent media channel configurations, the latter effectively providing a 375 GHz wide spectrum slot along with the approximation of the 69.4 GBd signal spectral shape. We call the representation of estimated GSNR over the OSaaS spectrum a GSNR profile. GSNR profile helps the wide-band OSaaS end users to evaluate the impact of the wavelength dependent performance variations on achievable throughput in the OSaaS spectrum. It demonstrates the severity of filtering at the media channel edges and any ripple/tilt in the OSaaS performance over the spectrum. In general, GSNR profiles captured with the smallest bandwidth and frequency increment provided by the PLT unit have the highest granularity for intra-spectrum performance estimations. This is due to the fact that wide-band PLT configurations may not distinguish any dips in the spectrum caused by adjacent concatenated media channels. Generally wider frequency step may not capture the smaller performance changes in the spectrum. This is illustrated with the coarse sample set of the DP-P-16QAM modulation format data set on Fig. 8. GSNR profile captured with smaller frequency increments, than any modern coherent signal bandwidth enables OSaaS customers to decide the best possible transceiver configurations for any part of the spectrum, to maximize the OSaaS throughput.

According to our measurements, all three modulation formats predict a similar GSNR performance with a maximum deviation of 0.25 dB in $GSNR_{est,link}$ for the 400 GHz spectrum configuration, and up to 0.53 dB $GSNR_{est,link}$ difference between the captured $GSNR_{est,link}$ datapoints for the coarse sample set of the 5x75 GHz configured OSaaS. Higher deviation in results in this case can be attributed to a time dependent variation in the network performance, as the samples in this case were collected on different days. Based on the GSNR profiles, 400 GHz single MC and 5x75 GHz adjacent MC OSaaS configurations have a different wavelength dependent performance that is caused by the different intra-OSaaS equalization schemes.

The most evident is a 2.5-dB GSNR tilt over the 400 GHz wide MC configuration that leads the DP-16QAM probe configuration, requiring generally higher GSNR to lack sufficient margin to work at the lower part of the spectrum slot. Although this difference can be reduced to 0.3 dB GSNR tilt when operating 5x75 GHz network media channel configuration, it reveals that the signal degradation caused by underlying infrastructure cannot fully be diminished with equalization. This can be accounted for the Raman-enabled amplification scheme of the LH route, creating a spectral ripple in the amplified C-band region with peak performances following the Raman pump locations and valleys in the between areas. Although the equalization at the ROADM nodes helps to balance out this difference, the underlying effects from the longest hops on the infrastructure are still detectable in estimated $GSNR_{est,link}$. Based on the longest span examples, the signals travelling in the valley part of the spectrum are more prone to linear OSNR depletion in the end of the links due to reduced effectiveness of the Raman amplification in the valley areas of the spectrum, resulting in overall reduced power levels and OSNR. The signals travelling at the peak parts of the spectrum are more prone to nonlinear effects, as their power levels in the beginning of each new network span after the equalizing ROADM are generally higher than average. While the ripple effect may not be relevant for medium-haul distances, it does create a performance difference between different spectral locations on the longer spans.

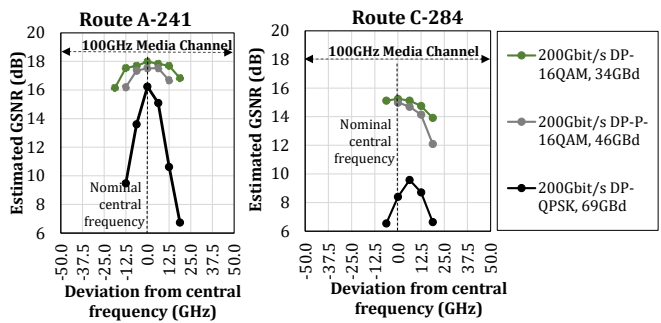

Fig. 7 Channel probing sweep results on a) route A and b) route C

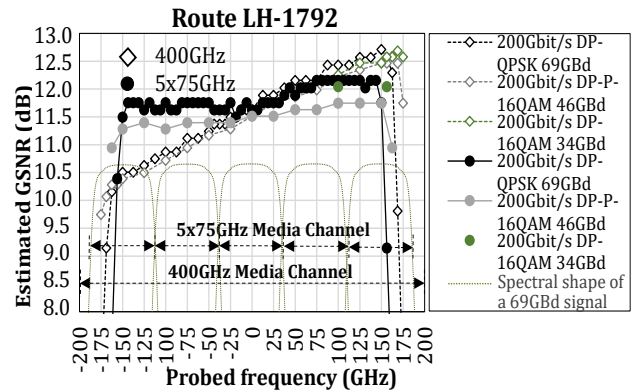

Fig. 8 Sweep results on LH-1792 with 400 GHz MC and 5x75 GHz MC

Based on the estimated GSNR and GSNR implementation margins at different parts of the spectrum on our example route, 300-Gbit/s 69-GBd DP-P-16QAM modulation could be used at the lower end of the spectrum, while 300-Gbit/s 52-GBd DP-16QAM signals are possible at the higher end. For the 5x75 GHz media channel configuration, regardless of the 0.3 dB Q-value difference over the spectrum service, the achieved GSNR allows only 300-Gbit/s 60-GBd DP-16QAM/DP-P-16QAM hybrid configuration to be used over the whole 5x75GHz spectrum. In addition, 200-Gbit/s 34-GBd DP-16QAM PLT configuration in 5x75 GHz media channel configuration only works for a few central frequencies over the tested spectral slot, experiencing post-FEC errors on the rest of the tested frequencies.

Although for the current link both configurations resulted in the same usable capacity carried within the tested wide-band OSaaS service, longer links may introduce a difference in total achievable bandwidth between two configurations. At any case, the GSNR performance information obtained by this measurement could be used for a power pre-emphasis at the transmitter to equalize the signal performance across the spectral slot and increase the service margins.

**C. Channel probing with power adjustments**

According to [36], the maximum channel performance is reached at an optimal power, where the amplified spontaneous emission noise power is twice the nonlinear noise power. In many production networks, per channel to-the-line launch powers are optimized per amplifier output, by dividing the maximum amplifier output power with the number of end-of-life channels. This is mostly a fixed figure used to commission all the links and is rarely adjusted according to the specific span length between two amplifier stations or even less likely, according to the total link length. And although the available simulation or planner tools enable to estimate the non-linear penalty per link, the opportunity is mostly used in the commissioning stage of the new network and forgotten in operation state. While the new, high-performing transceivers enable to overcome the small degradations or changes implemented after the initial commissioning, it may introduce changes in the optimal operation regime and reduce service margins.

One straight-forward way to detect the per channel optimum operation is to increase the individual channel powers to see when the top of the SNR vs. power (or "bell") curve is achieved. Unfortunately, within power limited networks, this could lead to amplifier saturation or performance degradation on neighbouring channels due to spectral hole burning effect. Therefore, a method that accounts for the maximum total power allowed by the spectrum slot is required.

This can be achieved by comparing the results received from symbol rate variable channel probing with constant PSD to the constant signal power measurements. Measurements with constant signal power use only the highest allowed total signal power for all tested symbol rates (effectively, increasing the PSD for the narrower signal formats). In both cases, the total power per probing activity stays equal or below the maximum allowed power per spectral slot and reference power used for the network equalization.

A selection of links on the route A in the regional-haul network were analysed by comparing the design-based PSD mode GSNR estimations to design based signal power mode GSNR estimations. The operation regime detection results are presented on Fig. 9. The solid lines are the probing results with single, constant signal power for all PLT configurations and dotted lines are the same PLT configurations tested with PSD adjusted power levels. Different shades/colours refer to different link lengths. The visible symbol rate dependent tilt on the GSNR estimations on the lines measured with constant PSD refer to filtering penalty from the route. As a reference, also one link from the long-haul network is included on the Fig. 9, presented with diamond marker shape. For each of the links, the reference PSD and reference total allowed signal power per carrier is at 69 GBd, where the measurement lines for the constant PSD and constant signal power meet.

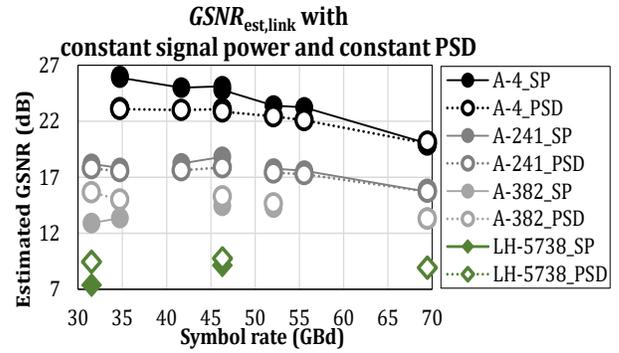

Fig. 9 Detection of the operation regime on different links

The operation regime of the specific configuration can be detected by comparing the estimated GSNR values for the constant PSD and constant signal power line. For the shorter links (black and dark gray lines), probing with constant signal powers provides better GSNR performance than probing with constant PSD. This means, the signals still do benefit from increased launch power, hence working in the linear regime. For the longest link in regional-haul network (light gray) and long-haul network (green), the constant signal power probing provides worse GSNR estimations, than the constant PSD probing. This indicates that, with constant signal power, the increased PSD of individual narrowband signals is too high, and the signal power is above the optimum between ASE noise and nonlinear distortions. This means, a reduction of the signal power at the specific OTSi should be considered.

It is worth mentioning that while the simple operation regime detection is sufficient in narrow band OSaaS configurations, wide-band OSaaS offerings may experience operation regime differences in different spectral parts of the OSaaS service. This has an increased probability in cases where high tilt is captured through the spectral sweep exercise. It is therefore recommended to carry out operation regime detection over the spectral range of the OSaaS service to characterize the spectral slot for highest usable power levels.

By detecting the operation regime, operators can adjust the to-the-line power levels per channel and per link to increase the service tolerance against random fluctuations in network performance that can lead to outages, when channels are operated at the BER threshold. Based on our example, it can be seen that while the $GSNR_{margin}$ can be improved up to +/- 2.5 dB for 31.5-GBd and 34.7-GBd signals depending on the link, the improvement on 46.3-GBd and 52.0-GBd signals is less, than 0.8 dB. However, this can be a great improvement when operated in low-margin regime.

The detected operation regime and possibility to increase to-the-line launch power levels only applies to narrow-band signals and does not impact the overall maximum allowed signal power per spectrum slot. Therefore, the power levels of the higher symbol rate signals can be only adjusted to lower levels, to avoid violations against contractually agreed highest total power per spectral slot. With verified filtering penalty, the power cut away by the filtering elements on higher symbol rate signals can be compensated by increasing the equalization threshold in the ROADM for the specific channel. However, verification of this opportunity is out of the scope of this paper, as it would require changes in the OLS parameters that spectrum end user cannot control.

## 5. FURTHER SYSTEM OPTIMIZATION POTENTIAL

In this section, we propose network optimization use-cases that focus on the throughput optimization potential based on the data acquired through the spectrum probing process.

### A. OLS optimization potential

Based on the results from spectrum QoT characterization with symbol rate variable channel probing in narrow-band OSaaS scenario, it is possible to evaluate the impact of the filtering penalty on the total achievable capacity on the network link. To illustrate this, we use the probing results for the regional-haul network from section 3-A. First, we plot the achievable throughput per channel and link based on $GSNR_{est,link}$ and $GSNR_{margin}$. These are illustrated with black bars on Fig. 10. Then, the potential throughput increase is estimated based on the PLT configuration that did not experience bandwidth limitation: 200-Gbit/s 34-GBd DP-16QAM or 100-Gbit/s 31-GBd DP-QPSK. These throughput calculations, indicating the achievable throughput gain from a system without any filtering penalty, are shown with green bars on Fig. 10. While only a small increase can be achieved due to moderate filtering effects on short distances, up to 100% increase can be achieved on links with additional filtering penalty from the DCG modules. For some links, getting rid of filtering elements would only bring performance improvement in service margin, but not a change in the maximum achievable line rate (link length 675 km). However, considering the average increase per channel, the total throughput increase would aggregate to 4-Tbit capacity gain, when implemented over the full C-band in a 40-channel system.

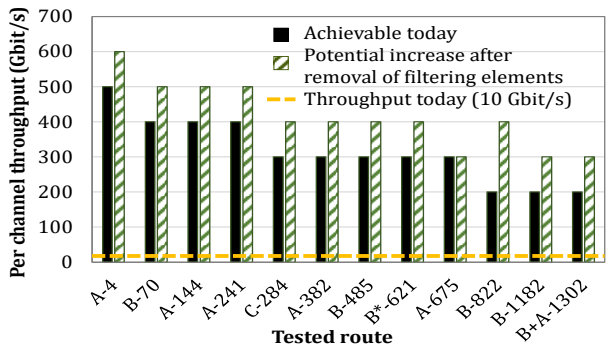

Fig. 10 Achievable and potential throughput estimations

This data can be used to make network upgrade decisions like adding colourless access structures. Caution must be taken when swapping DCG based dispersion modules to DCF based modules, as this could increase the nonlinear penalty on the link that might deplete the expected benefits from swapping the modules on the longer links (675-km link example). If possible, in the long run, going uncompensated and fully coherent also on the old legacy infrastructure links should be considered as an option instead.

### B Time-dependent optimization potential

While the primary usage scenario for the channel probing toolkit is to characterize the spectrum slot QoT performance in disaggregated networking scenarios, the initial spectrum characterization only covers the QoT performance at the time of the characterization. To overcome that, continuous channel probing utilizing already commissioned transceivers could be used to monitor the changes in OSaaS service performance. These changes could be caused by daily and yearly variations in the network, added crosstalk penalty from changed channel load in the neighbouring spectrum or other. Continuous channel probing enables spectrum end users to learn about the daily and yearly changes in the network performance and use the gained knowledge in running higher capacities during the network peak performance times. This can be useful in data-centre connectivity. According to our observations over the 10-month test-period, daily changes on a LH-3752 link during the summertime, impacted the estimated link GSNR on average 1.5 dB per 24 h period, while the performance fluctuations during the colder months reached at most 0.4 dB in amplitude per 24 hours. Since a better performance was observed during the colder night-time, users can optimize their higher-capacity demanding database back-ups to be run at night-time. On the observed link, 1.5 dB of GSNR difference would allow to switch from the default 200-Gbit/s 69-GBd DP-QPSK signal configuration to 200-Gbit/s 58-GBd DP-QPSK/DP-P-16QAM hybrid modulation and therefore reduce the required bandwidth from 75 GHz to 62.5 GHz per carrier, resulting in 2.56 Tbit/s capacity gain over C-band.

The benefits of continuous channel probing can also be used by the OLS operator. Probes operated inside the guard-band channels can be treated as a network monitoring point to perform OSaaS service policing in addition to general network performance monitoring and fault detection. Single characterized transceiver can be used to monitor multiple links. For this, an optical switch could be used to send the probing signal into different channel locations in multi-domain disaggregated network environments, providing ms-range switching time on already pre-provisioned lightpaths compared to reconfigurations in the ROADMs. These usage scenarios require additional research to provide solutions that reduce the investments costs per monitoring point.

## 6. CONCLUSIONS

In this paper, we have presented a practical toolkit based on the channel probing concept to characterize the QoT performance of the Optical Spectrum as a Service (OSaaS) in narrow-band and wide-band configurations. We have shown that the proposed methods and processes work over different network scenarios, utilizing both dispersion compensation free flex-grid systems and dispersion compensated fixed-grid filter-based systems. Symbol rate variable extended channel probing has been demonstrated to achieve a GNSR estimation accuracy of 0.05 dB for a wide-band OSaaS service compared to previously reported 0.7 dB and 0.32 dB for a narrow-band OSaaS service experiencing strong narrow band filtering. We have further presented the usefulness of spectral sweeping to capture the GSNR profile that could be used in OSaaS service contracts to demonstrate the service bandwidth and give insights to the OSaaS service performance and configuration. We have also demonstrated the usefulness of the operation regime detection in order to detect wavelength dependent performance differences within OSaaS, which allows to maximize service margins. Based on the performance data acquired from the proposed channel probing toolkit, spectrum users and OLS system operators can take fully informed decisions on how to leverage their spectral resources in the most efficient way without prior knowledge on the link set-up, channel transfer function or other impacting aspects, like operation regime, filtering penalty, neighbouring channel crosstalk, or other, directly contributing to GSNR of the service.

In our future research, we will investigate the impact of the different OSaaS configurations to GSNR profiles and compare it to other available metrics, capturing wavelength dependant performance relevant for OSaaS services.

**Funding Information.** The work has been partially funded by the Bundesministerium für Bildung und Forschung (BMBF) in the project OptiCON (grant 16KIS0989K).

**Acknowledgment**. We thank Tele2 Estonia for their continuous co-operation and help regarding the research on Optical Spectrum as a Service in Disaggregated Networks.